\documentclass[twocolumn,showpacs,preprintnumbers,amsmath,amssymb]{revtex4}

\usepackage{graphicx}
\usepackage{dcolumn}
\usepackage{bm}

\begin{document}

\preprint{APS/123-QED}

\title{Statistical theory of shot noise in quasi-1D Field Effect Transistors in the presence of electron-electron interaction}

\author{Alessandro Betti}
\affiliation{Dipartimento di Ingegneria dell'Informazione: Elettronica, Informatica, Telecomunicazioni, \\Universit\`a di Pisa, Via Caruso 16, 56122 Pisa, Italy.}
\author{Gianluca Fiori}%
\affiliation{Dipartimento di Ingegneria dell'Informazione: Elettronica, Informatica, Telecomunicazioni, \\Universit\`a di Pisa, Via Caruso 16, 56122 Pisa, Italy.}

\author{Giuseppe Iannaccone}
\affiliation{Dipartimento di Ingegneria dell'Informazione: Elettronica, Informatica, Telecomunicazioni, \\Universit\`a di Pisa, Via Caruso 16, 56122 Pisa, Italy.}

\begin{abstract}
We present an expression for the shot noise power spectral density in 
quasi-one dimensional conductors 
electrostatically controlled by a gate electrode, that includes 
the effects of Coulomb interaction and of Pauli exclusion among 
charge carriers. In this sense, our expression extends the well known 
Landauer-B\"uttiker noise formula
to include the effect of Coulomb interaction inducing fluctuations 
of the potential in the device region.
Our approach is based on evaluating the statistical properties of 
the scattering matrix and on a second-quantization many-body description. From a quantitative point of view, statistical properties are obtained by means 
of Monte Carlo simulations on an ensemble of different configurations of 
injected states, requiring the solution of the Poisson-Schr\"odinger 
equation on a three-dimensional grid, with the non-equilibrium Green's 
functions formalism. In a series of examples, we show  that failure to 
consider the effects of Coulomb interaction on noise leads to a gross 
overestimation of the noise spectrum of quasi-one dimensional devices.

\end{abstract}

\pacs{73.50.Td, 73.63.Nm}
\keywords{Shot noise, FETs, nanowire transistors, carbon nanotube 
transistors.}
\maketitle

\section{Introduction}
As quasi one-dimensional field-effect transistors (FETs), based for example
on Carbon Nanotubes (CNTs) or Silicon NanoWires (SNWs), are increasingly
investigated as a possible replacement for conventional planar FETs, it is 
important to achieve 
complete understanding of the properties of shot noise of one-dimensional 
conductors electrostatically controlled by a third (gate) electrode. 
Shot noise is particularly
sensitive to carrier-carrier interaction, that in turn can be particularly 
significant in one-dimensional
nanoscale conductors, where electrons are few and screening is 
limited~\cite{Land}. 

Low frequency $1/f$ noise in quasi one-dimensional conductors has been the 
subject of interest for several authors~\cite{YMLin,JAppenzeller,JTersoff}, 
whereas few experimental papers on shot noise have recently been 
published~\cite{PRoche,Hermann}.
 
Due to the small amount of 
mobile charge in nanoscale one-dimensional FETs, even in strong inversion, 
drain 
current fluctuations can heavily affect device electrical behavior. 
Of course, noise is an unavoidable and undesirable feature of electron devices,
and its effect must be minimized or kept within tolerable levels for the operation of 
electronic circuits. From a more fundamental point of view, it is also a rich 
source of information on electron-electron interaction, 
which cannot be obtained from DC or AC electrical characteristics.

The main sources of noise are injection from the contacts into the 
device region, through the
random occupation of states around the Fermi energy at the contacts, and 
partial transmission of
electrons through the conductor, which gives rise to the so called 
partition noise. The main types of interaction
that have a clear effect on noise are Pauli exclusion, which reduces 
fluctuations of the rate of injected electrons
by limiting the occupancy of injected states, and Coulomb repulsions among 
electrons, which is the cause of 
fluctuations of the potential in the device region, that often suppress, 
but sometimes enhance the effect of fluctuations in the rate of injected 
electrons. 

The combined effect of Pauli exclusion and Coulomb repulsion on shot 
noise has been investigated in the case of  ballistic 
double gate MOSFETs~\cite{YNaveh}, in planar MOSFETs~\cite{ik1_2} and 
in resonant tunneling diodes~\cite{ik1,ik1PRB,BlantBut}. 
There are still few attempts~\cite{Bul1} to a complete quantitative 
understanding of shot noise in ballistic CNT- and SNW-FETs. 
Indeed, when addressing a resonant tunneling diode one can
usually adopt an approach that exploits the fact that the two opaque barriers break
the device in three loosely coupled regions (the two contacts and the well), among which
transitions can described by Fermi golden rule, 
as has been done in Refs. \cite{ik1,ik1PRB,BlantBut}. This is not possible
in the case of a transistor, where coupling between the channel and the contacts is very good.

Another important issue is represented by the fact that the widely known 
Landauer-B\"uttiker's noise formula~\cite{MBut2,TMartin}, does not 
take into account the effect of Coulomb interaction on shot noise through potential fluctuations.
Indeed, recent experiments on shot noise in CNT-based Fabry Perot 
interferometers~\cite{Hermann} show that in some bias conditions 
many-body corrections might be needed to explain the observed noise 
suppression.  
Other experiments show that at low temperature suspended 
ropes of single-wall carbon nanotubes 
of length 0.4 $\mu$m exhibit a significant 
suppression of current fluctuations by a factor smaller than 1/100 
compared to full shot noise~\cite{PRoche}. However, this 
experimental result is not supported by a convincing interpretation, since possible explanations 
extend from ballistic transport in a small number of tubes within a rope, 
to diffusive transport in a substantial fraction of the CNTs.

In this work, we present an expression for the shot noise power 
spectral density of ballistic quasi-one dimensional channels based on a statistical approach 
relying on quantities obtained from Monte Carlo (MC) 
simulations of randomly injected electrons from the reservoirs. 
The expression is derived within the second quantization 
formalism, and simulations are based on the self-consistent solution of the 
3D Poisson and Schr\"odinger equations, within the non-equilibrium Green's 
function (NEGF) formalism~\cite{Datta}.

Our proposed expression generalizes the Landauer-B\"uttiker's noise formula 
including the effects of Coulomb interaction, that is significant for a 
large class of devices, and in particular for one-dimensional conductors.

\section{Theory}

According to Milatz's theorem~\cite{Ziel}, the 
power spectral 
density of the noise current in the zero frequency limit can be written as 
$ S(0) =$ lim$_{1/\nu \rightarrow  \infty} \left[2/\nu \cdot var(I)\right] $, 
where  $\nu$ is the injection rate of a carrier from a contact and 
$var(I)$ is the variance of the current. 
According to Ref.\cite{MButtiker}, $\nu$ can be expressed as  
$ \nu = \Delta E/(2 \pi \hbar) $ where $\Delta E$ is the energy 
discretization step, i.e. the minimum energy separation 
between injected states. 
Indeed, the contribution to the current of a transverse mode in the energy 
interval $\Delta E$ can be expressed in the zero temperature limit by the 
Landauer-B\"uttiker formula as 
$ \langle dI \rangle= e/(2\pi\hbar) \, \Delta E $. On the other 
hand $ \langle dI \rangle = e \nu$, from which 
$ \nu = \Delta E/(2 \pi \hbar) $ derives. 
Finally, the power spectral density of shot noise at zero frequency can be 
expressed as:
\begin{eqnarray} \label{eqn:noisepower}
S(0) &=& \lim_{\nu \rightarrow  0}\frac{2}{\nu} \, var(I)= \lim_{\Delta E \rightarrow  0} 4 \pi \hbar \frac{var(I)}{\Delta E} \, .
\end{eqnarray}
The variance of the current can be derived 
by means of the second quantization formalism, which allows a concise 
treatment of the many-electron problem.

\begin{figure} [tbp]
\begin{center}
\includegraphics[width=8.0cm]{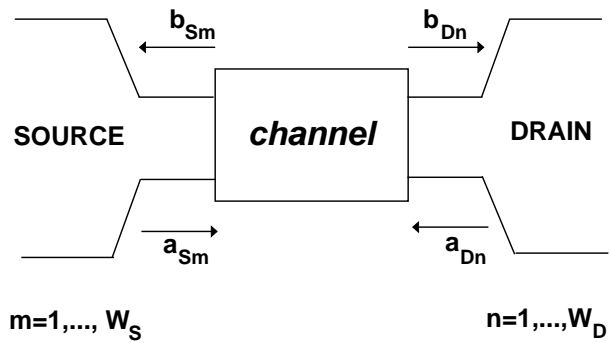}
\end{center}
\caption{Annihilation operators for ingoing ($a_{Sm}$, $a_{Dn}$) and 
outgoing electron states ($b_{Sm}$, $b_{Dn}$) in a two terminal 
scattering problem ($m=1,...,W_S$; $n=1,...,W_D$).}
\label{fig:scattering}
\end{figure}

Let us consider a mesoscopic conductor connected to two 
reservoirs [source (S) and drain (D)], where electron states are populated 
according to their Fermi occupation factors (Fig.~\ref{fig:scattering}).
For simplicity, we assume that the conductor is sufficiently short as to completely neglect inelastic scattering events. Thermalization occurs only in the reservoirs. 
At zero magnetic field and far from the interacting channel, 
the time-dependent current operator at the source can be expressed 
as the difference between 
the occupation number of carriers moving inward 
($N_{Sm}^+$) and outward ($N_{Sm}^-$) in each quantum channel $m$~\cite{MBut2}:
\begin{eqnarray} \label{eqn:current}
I(t) \! \! &=& \! \!\frac{e}{h} \sum_{m \in S} \int d\!E \left[N_{Sm}^+(E,t) - N_{Sm}^-(E,t)\right] \, ,
\end{eqnarray}
where 
\begin{eqnarray} \label{eqn:n+}
N_{Sm}^+(E,t) \!\! \! \! &=& \! \!\! \!\int d(\hbar \omega) a_{Sm}^+(E)\,a_{Sm}(E+\hbar \omega)\, e^{-i\omega t} \, , \nonumber \\
N_{Sm}^-(E,t) \!\! \! \! &=& \! \!\! \!\int d(\hbar \omega) b_{Sm}^+(E)\,b_{Sm}(E+\hbar \omega)\, e^{-i\omega t} \, .
\end{eqnarray}
The introduced operators $a_{Sm}^{\dagger}(E)$ and $a_{Sm}(E)$ create and 
annihilate, respectively, incident electrons in the 
source lead with total energy $E$ in the channel $m$ 
(Fig.~\ref{fig:scattering}). 
In the same way, 
the creation $b_{Sm}^{\dagger}(E)$ and annihilation $b_{Sm}(E)$ operators 
refer to electrons in the source contact for outgoing states. 
The channel index $m$ runs over all the transverse modes and 
different spin orientations.

The operators $a$ and $b$ are related via an unitary transformation 
($n=1,...,W_S$)~\cite{MBut2}:
\begin{eqnarray} \label{eqn:b(a)} 
b_{Sn}(E)\!\!\!&=&\!\!\!\sum_{m = 1}^{W_S} \mathbf{r}_{nm}(E) a_{Sm}(E)\!+\!\!\! \sum_{m = 1}^{W_D} \mathbf{t'}_{nm}(E) a_{Dm}(E) \, ,  \nonumber \\
\end{eqnarray}
where $W_S$ and $W_D$ represent the number of quantum channels in the source 
and drain leads, respectively, while the blocks $\mathbf{r}$ 
(size $W_S \times W_S$) and $\mathbf{t'}$ (size $W_S \times W_D$), 
describe electron reflection at the source ($\mathbf{r}$) and 
transmission from drain to source ($\mathbf{t'}$) and 
are included in the scattering matrix $\mathbf{s}$ as~\cite{Datta1}:

\begin{eqnarray} \label{eqn:s}
\mathbf{s}= \left(
\begin{array}{cc}
 \mathbf{r} & \mathbf{t'}   \\
 \mathbf{t} & \mathbf{r'}   \\     
\end{array}
\right) \, .
\end{eqnarray} 
The dimensions of $\mathbf{s}$ are $(W_S+W_D)\times(W_S+W_D)$. 
Blocks $\mathbf{t}$ and $\mathbf{r'}$ in Eq.~(\ref{eqn:s}) are related to 
source-to-drain 
transmission and reflection back to the drain, respectively. 
In the following, time dependence will be neglected, since we are interested 
to the zero frequency case.

If we denote with $\mid \!\!\sigma \rangle$ a many-particle (antisymmetrical) 
state, 
the occupation number in the reservoir $\alpha$ in the channel $m$ can be 
expressed as $\sigma_{\alpha m}(E)= \langle a_{\alpha m}^{\dagger}(E)a_{\alpha m}(E)\rangle_{\sigma}$. Pauli exclusion principle does not allow two 
electrons to occupy the same spin orbital, therefore $\sigma_{\alpha m}(E)$ can 
be either 0 or 1. In addition, since fluctuations of the potential profile along the channel due to Coulomb 
interaction between randomly injected carriers affect the transmission of 
electrons, the scattering matrix elements have to depend on the occupation 
numbers of all states in both reservoirs: $\mathbf{s}(E)= \mathbf{s}\left[\sigma_{S 1}(E),\sigma_{S 2}(E),\cdots,\sigma_{D 1}(E),
\sigma_{D 2}(E),\cdots\right]$.
Let us stress the fact that, as pointed out in 
Ref.~\cite{MBut2}, whenever a 
finite channel is connected to semi-infinite leads, the channel can be 
considered as a small perturbation to the equilibrium regime of the contacts, 
and independent random statistics can be used for both reservoirs.

According to Ref.~\cite{MBut2}, current fluctuations can be evaluated by 
introducing an ensemble of 
many electrons states $\{\mid  \!\! \sigma_1 \rangle,
\mid   \!\! \sigma_2 \rangle,$ $\mid  \! \sigma_3 \rangle,\cdots,
\mid  \! \sigma_N \rangle\}$ and by weighting each state properly, i.e. by 
finding its statistical average, denoted by $\langle \, \rangle_s$. 
Each reservoir $\alpha \,(\alpha=S,D)$ is assumed to be at thermal 
equilibrium, so that its average occupancy can be described 
by the Fermi-Dirac statistics $f_{\alpha}$. 
As a consequence, 
the statistical average of $\sigma_{\alpha m}(E)$ reads~\cite{MBut2}: 
\begin{eqnarray}\label{eqn:relat} 
\langle \sigma_{\alpha m}(E) \rangle_s= \,\langle \langle a_{\alpha m}^{\dagger}(E) a_{\alpha m}(E)\rangle_{\sigma}\rangle_s = \, f_{\alpha}(E) \, .
\end{eqnarray} 
Neglecting correlations between the occupation numbers of 
the same quantum channel at different energies, or between different channels 
at the same energy, we obtain~\cite{MBut2}: 
\begin{eqnarray} \label{eqn:relatparz}
\langle \sigma_{\alpha m}(E) \sigma_{\beta n}(E') \rangle_s \!\!\!&=&\!\!\!  f_{\alpha}(E)f_{\beta}(E')  
\end{eqnarray} 
for $\alpha\neq \beta$ or $m \neq n$ or $E \neq E'$. 
Including Eq.~(\ref{eqn:relat}) in Eq.~(\ref{eqn:relatparz}) and exploiting 
the relation $\sigma_{\alpha m}(E)^2=\sigma_{\alpha m}(E)$, the 
average of the product of two occupation numbers can be expressed as: 
\begin{eqnarray} 
\langle \sigma_{\alpha m}(E) \sigma_{\beta n}(E') \rangle_s \!\!\!&=&\!\!\!  f_{\alpha}(E)f_{\beta}(E') + \delta(E\!-\!E')\delta_{\alpha \beta} \delta_{m n} \nonumber \\
&& \left[f_{\alpha}(E)\!-\!\!f_{\alpha}(E)f_{\beta}(E')\right] \,, 
\end{eqnarray}
where $\delta(E-E')$, $\delta_{\alpha \beta}$, $\delta_{m n}$ are Kronecker 
delta functions.

In order to compute the average current 
along the channel and the power spectral density of the current fluctuations, 
we need to write the expectation values of the products of two and four 
operators~\cite{MBut2}:
\begin{eqnarray} \label{eqn:relation1}
&&\langle a_{\alpha m}^{\dagger}(E) a_{\beta n}(E')\rangle_{\sigma}= \delta(E-E') \delta_{\alpha \beta} \delta_{m n} \sigma_{\alpha m}(E) \, ; \\
\nonumber \end{eqnarray}
\vspace{-1.3cm}
\begin{eqnarray} \label{eqn:relation2}
&&\langle a_{\alpha m}^{\dagger}(E) a_{\beta n}(E') a_{\gamma k}^{\dagger}(E'') a_{\delta l}(E''')\rangle_{\sigma}= 
\nonumber \\
&& \delta(E\!-\!E''') \delta(E'\!-\!E'') \delta_{\alpha \delta} \delta_{m l} \delta_{\beta \gamma} \delta_{n k} \sigma_{\alpha m}(E) \left[1-\sigma_{\gamma k}(E'') \right] 
 \nonumber \\
&& +\, \delta(E\!-\!E') \delta(E''\!-\!E''') \delta_{\alpha \beta} \delta_{n m} \delta_{\gamma \delta} \delta_{k l} \sigma_{\alpha m}(E) \sigma_{\gamma k}(E'')    \\
\nonumber \end{eqnarray} 
where the first contribution in Eq.~(\ref{eqn:relation2}) refers to 
exchange pairing ($\alpha=\delta$, $\beta=\gamma$, $m=l$, $n=k$), while the 
second to normal pairing 
($\alpha=\beta$, $\gamma=\delta$, $m=n$, $k=l$)~\cite{MBut2}. 
For the sake of simplicity, in the following we denote the expectation 
$\langle \langle \,\,\, \rangle_{\sigma}\rangle_{s}$ as 
$\langle \,\,\, \rangle$. 

By means of Eqs.~(\ref{eqn:b(a)}) and~(\ref{eqn:relation1}) 
the average current reads:
\begin{eqnarray} \label{eqn:meancurrent}
\langle I \rangle \! \!\!&=&\!\! 
 \frac{e}{h}\! \int \!d\!E \!\left\{\sum_{n \in S} \langle \left[\mathbf{t^{\dagger} t}  \right]_{nn} \sigma_{S n}\rangle_s -\!\! \sum_{k \in D} \langle \left[\mathbf{t'^{\dagger}t'} \right]_{kk} \sigma_{D k}  \rangle_s \!\right\} \nonumber \\
&=&\!\!\! \frac{e}{h}\! \int \!\!d\!E \!\left\{\sum_{n \in S} \langle \left[\mathbf{\tilde{t} } \right]_{S;nn} \sigma_{S n}\rangle_s -\!\! \sum_{k \in D} \langle \left[\mathbf{\tilde{t}} \right]_{D;kk} \sigma_{D k}  \rangle_s \!\!\right\} ,\nonumber \\
\end{eqnarray}
where $\left[\mathbf{\tilde{t}}\right]_{\alpha;lp} \equiv \left[\mathbf{t^{\dagger} t}\right]_{lp}$ if $\alpha=S$ and 
$\left[\mathbf{t'^{\dagger} t'}\right]_{lp}$ if $\alpha=D$ 
($l,p \in \alpha $). 
The unitarity of the matrix $\mathbf{s}$ has also been exploited, from which 
the relation $\mathbf{r^{\dagger} r} + \mathbf{t^{\dagger} t}= \, \mathbf{1}$ follows.
It is easy to show that for a non-interacting channel, i.e. when 
occupancy of injected states does not affect 
transmission and reflection probabilities, 
Eq.~(\ref{eqn:meancurrent}) reduces to the two-terminal 
Landauer's formula~\cite{RLandauer}. 

In general, we can observe that for an interacting channel 
Eq.~(\ref{eqn:meancurrent}) 
provides a different result with respect to Landauer's formula, because 
fluctuation of transmission probabilities induced by random injection in 
the device,
is responsible for rectification of the current. The effect is often very 
small,but not always~\cite{ABettiTED}.  
However, it cannot be captured by Landauer's formula, as other
many-particle processes affecting device transport 
properties~\cite{NSai,Vignale}. 

The mean squared current reads:
\begin{eqnarray} 
\langle I^2 \rangle \!\!&=&\!\!\! \left(\frac{e}{h}\right)^2  \! \! \int  \!d\!E  \!\!\int  \!d\!E' \! \sum_{m,n \in S}\!  \left\{ \langle  N_{Sm}^+(E) N_{Sn}^+(E')\rangle \right. \nonumber \\
&-&  \langle   N_{Sm}^+(E)  N_{Sn}^-(E') \rangle   - \langle  N_{Sm}^-(E) N_{Sn}^+(E')\rangle\!\!   \nonumber \\ 
&+&  \left. \langle   N_{Sm}^-(E)  N_{Sn}^-(E') \rangle\! \right\}\nonumber \\ 
&=&  F_{++}+F_{+-}+F_{-+}+F_{--} \, .
\end{eqnarray}
This expression consists of four terms, related to states at the source 
contacts, that can be evaluated by means 
of Eqs.~(\ref{eqn:relation1}) and~(\ref{eqn:relation2}): the first one 
($F_{++}$) represents the 
correlation of fluctuations in two ingoing
streams, the second and 
the third ones ($F_{+-}$, $F_{-+}$) describe the correlations of the 
fluctuations of the ingoing 
and outgoing streams, the fourth one ($F_{--}$) refers to two 
outgoing streams.

The first term $F_{++}$ can be expressed as:
\begin{eqnarray} \label{eqn:F++}
F_{++}= 
\left(\frac{e}{h}\right)^2 \!\!\int \!\!d\!E \!\int \!\!d\!E' \!\!\!\sum_{m,n \in S} \langle \sigma_{Sm}(E)\sigma_{Sn}(E')\rangle_s \, ,
\end{eqnarray}
since $\langle \sigma_{Sm}^2(E) \rangle_s= \, \langle \sigma_{Sm}(E)\rangle_s= \, f_S(E)$ $\forall m \in S$. 
Correlations between ingoing states are established through the statistical 
expectation values of each couple of occupancies of states injected from the 
source.

The second contribution $F_{+-}$ reads:
\begin{eqnarray} \label{eqn:F+-} 
F_{+-}\!\!\!\!&=&\!\!\!-\left(\frac{e}{h}\right)^2 \!\!\int \!d\!E \!\!\int \!d\!E'\! \left\{\!\sum_{m,l \in S} \!\!\langle \left(1- \left[\mathbf{\tilde{t}} (E') \right]_{S;ll}\right)\! \sigma_{Sm}(E)  \right. \nonumber \\
&& \sigma_{Sl}(E')\rangle_s\! +\!\!\!\!\left. \sum_{m \in S}\sum_{k \in D} \langle \left[\mathbf{\tilde{t}} (E') \right]_{D;kk} \!\sigma_{Sm}(E) \sigma_{Dk}(E') \rangle_s \!\right\} , \nonumber \\
\end{eqnarray}
since $ \sigma_{\alpha l}^2(E)= \sigma_{\alpha l}(E) \,\,\forall  l\in \alpha$ $(\alpha=S,D)$, due to the Pauli exclusion principle. 
In Eq.~(\ref{eqn:F+-}) correlations between ingoing and outgoing states 
are obtained by summing on each 
statistical average of the product of two occupation numbers of 
injected states, 
weighted with the reflection ($1- \left[\mathbf{\tilde{t}} (E')\right]_{S;ll}=\left[\mathbf{r^{\dagger}r}(E')\right]_{ll}$) or transmission probability ($\left[\mathbf{\tilde{t}} (E')\right]_{D;kk}$) of outgoing channels.  

By exploiting the anticommutation relations of the fermionic operators $a$, 
it is simple to demonstrate that the third term $F_{-+}$ is identical to 
$F_{+-}$. Indeed:
\begin{eqnarray} 
\langle  N_{Sm}^-(E) N_{Sn}^+(E')\rangle 
= \langle  N_{Sn}^+(E') N_{Sm}^-(E) \rangle \, .
\end{eqnarray}
Finally, the fourth term $F_{--}$ reads:
\begin{eqnarray} \label{eqn:F--}
F_{--}\!\!\!&=&\!\!\!   \left(\frac{e}{h}\right)^2 \!\Delta E\!\int \!d\!E  \!\sum _{\alpha=S,D} \sum_{l \in \alpha} \langle  \left[\mathbf{\tilde{t}}\right]_{\alpha;ll}\left(1\!-\!\left[\mathbf{\tilde{t}}\right]_{\alpha;ll}\right)\sigma_{\alpha l}\rangle_s \nonumber \\
&-&  \!\! \left(\frac{e}{h}\right)^2 \!\!\Delta E \!\int \!d\!E \! \sum _{\alpha=S,D} \!\! \sum_{
\begin{array}{c}
\scriptstyle l,p \in \alpha \\
\scriptstyle l \neq p \\
\end{array}} 
\!\!\langle \left[\mathbf{\tilde{t}}\right]_{\alpha;l p} \left[\mathbf{\tilde{t}}\right]_{\alpha;p l} \sigma_{\alpha l} \sigma_{\alpha p}\rangle_s \nonumber \\
&-&\!\!\! 2 \left(\frac{e}{h}\right)^2 \!\!\Delta E\!\!  \int \!\!d\!E \!\sum_{k \in D} \sum_{p \in S}  \langle \left[\mathbf{t'^{\dagger}r}\right]_{kp} \left[\mathbf{r^{\dagger}t'}\right]_{pk}  \sigma_{Dk} \sigma_{Sp} \rangle_s  \nonumber \\  
&+&\!\!\! \langle \left[\frac{e}{h} \!\int \!d\!E \left( \sum_{l \in S} \left[\mathbf{\tilde{t}}\right]_{S;l l}  \sigma_{Sl}\! - \!\!\sum_{k \in D} \left[\mathbf{\tilde{t}}\right]_{D;k k}  \sigma_{Dk} \right) \right]^2 \!\rangle_s \nonumber \\
&+& \!\!\!2 \left(\frac{e}{h}\right)^2 \!\!\!\!\int \!\!d\!E  \!\!\! \int \!\!d\!E' \!\sum_{l \in S}  \!\sum_{k \in D}  \langle \left[\mathbf{\tilde{t}}(E')\right]_{D;kk} \!\sigma_{S l}(E)\sigma_{Dk}(E') \rangle_s \nonumber \\
&-& \!\!\!2 \left(\frac{e}{h}\right)^2 \!\int \!\!d\!E  \! \int \!d\!E' \sum_{l,p \in S}  \langle \left[\mathbf{\tilde{t}}(E)\right]_{S;ll} \sigma_{S l}(E)\sigma_{Sp}(E') \rangle_s \nonumber \\
&+& \!\! \left(\frac{e}{h}\right)^2 \!\!\int \!d\!E  \! \!\int \!d\!E' \sum_{l,p \in S}  \  \langle \sigma_{S l}(E)\sigma_{Sp}(E') \rangle_s \, .
\end{eqnarray}
Equation~(\ref{eqn:F--}) contains all correlations between outgoing 
electron states in the source lead, where outgoing carriers at the source can be either
reflected carriers incident from $S$ or transmitted carriers injected 
from $D$. 
By means of the Eqs.~(\ref{eqn:F++}), (\ref{eqn:F+-}) and 
(\ref{eqn:F--}), we find the mean squared current:
\begin{eqnarray} \label{eqn:meansquaredI}
\langle I^2 \rangle \!\!\!&=&\!\!\!   \left(\frac{e}{h}\right)^2 \!\!\Delta E\!\int \!d\!E  \!\!\sum _{\alpha=S,D} \sum_{l \in \alpha} \langle  \left[\mathbf{\tilde{t}}\right]_{\alpha;ll}\left(1\!-\!\left[\mathbf{\tilde{t}}\right]_{\alpha;ll}\right)\sigma_{\alpha l}\rangle_s \nonumber \\
&-&  \!\! \left(\frac{e}{h}\right)^2 \!\Delta E\int \!d\!E \! \!\sum _{\alpha=S,D}  \sum_{
\begin{array}{c}
\scriptstyle l,p \in \alpha \\
\scriptstyle l \neq p \\
\end{array}} 
\langle \left[\mathbf{\tilde{t}}\right]_{\alpha; l p} \left[\mathbf{\tilde{t}}\right]_{\alpha;p l} \sigma_{\alpha l} \sigma_{\alpha p}\rangle_s \nonumber \\
&-&\!\!\! 2 \left(\frac{e}{h}\right)^2 \!\!\Delta E \!\! \int \!\!d\!E \!\!\sum_{k \in D} \sum_{p \in S}  \langle \left[\mathbf{t'^{\dagger}r}\right]_{kp} \left[\mathbf{r^{\dagger}t'}\right]_{pk}  \sigma_{Dk} \sigma_{Sp} \rangle_s  \nonumber \\  
&+&\!\!\! \langle \left[\frac{e}{h} \!\int \!d\!E \left( \sum_{l \in S} \left[\mathbf{\tilde{t}}\right]_{S;l l}  \sigma_{Sl}\! - \!\!\sum_{k \in D} \left[\mathbf{\tilde{t}}\right]_{D;k k}  \sigma_{Dk} \right) \right]^2 \!\rangle_s \, .\nonumber \\
\end{eqnarray}
Finally, from Eqs.~(\ref{eqn:noisepower}),~(\ref{eqn:meancurrent}) 
and~(\ref{eqn:meansquaredI}) the noise power spectrum can be 
expressed as:
\begin{eqnarray} \label{eqn:variance}
S(0)\!\!\! &=&\!\!\! \left(\frac{e^2}{\pi \hbar}\right) \!\!\int \!\! d\!E \!\! \sum_{\alpha = S,D} \sum_{l \in \alpha} \langle  \left[\mathbf{\tilde{t}}\right]_{\alpha;ll}\left(1-\left[\mathbf{\tilde{t}}\right]_{\alpha;ll}\right) \! \sigma_{\alpha l}\rangle_s \nonumber \\
&-&\!\!\!\! \left(\frac{e^2}{\pi \hbar}\right)  \!\!\int \!\!d\!E \! \sum_{\alpha = S,D} \!
\sum_{
\begin{array}{c}
\scriptstyle l,p \in \alpha \\
\scriptstyle l \neq p \\
\end{array}} 
\langle \left[\mathbf{\tilde{t}}\right]_{\alpha;l p} \left[\mathbf{\tilde{t}}\right]_{\alpha;p l} \sigma_{\alpha l} \sigma_{\alpha p}\rangle_s \nonumber \\
&-& \!\!\!2  \left(\frac{e^2}{\pi \hbar}\right) \!\!\int \!\!d\!E \!\sum_{k \in D} \sum_{p \in S} \!\langle \left[\mathbf{t'^{\dagger}r}\right]_{kp} \left[\mathbf{r^{\dagger}t'}\right]_{pk}  \sigma_{Dk} \sigma_{Sp} \rangle_s  \nonumber \\
&+& \!\!\!\!\frac{4 \pi \hbar}{\Delta E} var \! \left\{\! \frac{e}{h}\! \int \!\!d\!E \!\left(\sum_{n \in S} \!\! \left[\mathbf{\tilde{t}} \right]_{S;nn}\! \sigma_{S n}  \! 
- \!\!\! \sum_{k \in D} \!\! \left[\mathbf{\tilde{t}} \right]_{D;kk}\! \sigma_{D k} \! \right)\!\right\} . \nonumber \\ 
\end{eqnarray} 
Equation~(\ref{eqn:variance}) is the main theoretical result of this work: 
the power spectral density of the noise current is expressed in terms of  
transmission ($\mathbf{t}$, $\mathbf{t'}$), reflection ($\mathbf{r}$) 
amplitude matrices, and properties of the leads, such as 
random occupation numbers of injected states. 
Let us point out that, although our derivation starts from 
Eq.~(\ref{eqn:current}), which 
is valid only far from the mesoscopic interacting sample, 
Eq.~(\ref{eqn:variance}) allows to take into account both Pauli and Coulomb 
interactions through the dependence of $\mathbf{t}$, $\mathbf{t'}$ and 
$\mathbf{r}$ on actually injected states. Let us note that we go beyond 
the Hartree approximation by considering different 
random configuration of injected electron states for different many-particle 
systems.

There is a crucial difference with respect to Landauer-B\"uttiker's 
formula, since Eq.~(\ref{eqn:variance}) enables 
to consider fluctuations in time of the potential profile along the channel 
induced by the electrostatic repulsion between randomly injected electrons 
from the leads. 
Essentially, for each random configuration of injected states from both 
reservoirs, we consider a \begin{it} snapshot \end{it} of device 
operation at a different time instant.
All statistical properties --- in the limit of zero frequency --- 
can be obtained by considering a sufficient ensemble of snapshots.

Let us discuss some physical limits of interest. 
First, we consider the case of zero temperature. 
In such condition the Fermi factor for populating electron states 
in the reservoirs is either 0 or 1, and all snapshots are identical, so
the fourth term in Eq.~(\ref{eqn:variance}) disappears. 
In addition, we can remove the statistical 
averaging in Eq.~(\ref{eqn:variance}) and the first three terms lead to the 
following expression of the noise power spectrum: 
\begin{eqnarray} \label{eqn:zerotemperature}
S(0)=  \frac{2\,e^2}{\pi\hbar} \!\int_{E_{FD}}^{E_{FS}} \!\! d\!E  \left( \mathrm{Tr}\left[\mathbf{t^\dagger t} \right]\!-\!\mathrm{Tr}\left[\mathbf{t^\dagger t t^\dagger t}\right]\!\right) \, ,
\end{eqnarray} 
where $E_{FS}$ and $E_{FD}$ are the Fermi energies of the source and drain 
contacts, respectively. 
Such terms can be identified with partition noise (PN) contribution.
More in detail, the first term of Eq.~(\ref{eqn:variance}) 
is associated to the 
quantum uncertainty of whether an electron injected in the mode $l$ 
from the reservoir $\alpha$ is transmitted through or reflected 
by the barrier. 

The second term of Eq.~(\ref{eqn:variance}) contains instead ($l \neq p$):
\begin{eqnarray} \label{eqn:coupl1}
\left[\mathbf{t^{\dagger}t}\right]_{l p}\left[\mathbf{t^{\dagger}t}\right]_{p l}= 
\sum_{k,q \in D} \mathbf{t}^{*}_{k l} \mathbf{t}_{k p} \mathbf{t}^{*}_{q p} \mathbf{t}_{q l} \, .
\end{eqnarray} 
Each term of the sum can be interpreted as the 
coupling between a transmission event from channel $p\in S$ into channel 
$k\in D$ and from channel $l\in S$ into channel $q\in D$: such a coupling 
is due to time-reversed transmissions from $k$ into 
$l$ and from $q$ into $p$.

In the same way, the third term of Eq.~(\ref{eqn:variance}) contains 
\begin{eqnarray} \label{eqn:coupl2}
\left[\mathbf{t'^{\dagger}r}\right]_{k p}\left[\mathbf{r^{\dagger}t'}\right]_{p k}= 
\sum_{l,n \in S} \mathbf{t}^{*}_{k l} \mathbf{r}_{l p} \mathbf{r}^{*}_{n p} \mathbf{t}_{k n} \, ,
\end{eqnarray} 
that represents the coupling between carriers transmitted from $n\in S$ into 
$k \in D$ and reflected from $p \in S$ into $l \in S$. 
The second and third terms provide insights on exchange 
effects. Indeed, in such terms, contributions with 
$k \neq q$ and $l \neq n$, respectively, are 
complex and they represent exchange interference effects (fourth-order 
interference effects) in the 
many-particle wave-function due to the quantum-mechanical 
impossibility to distinguish 
identical carriers~\cite{MButtiker}. 
In the Results section, we will be concerned with identical reservoirs, i.e. 
identical injected modes from the contacts. In this case the diagonal terms 
of the partition noise (first term and part of the third term in 
Eq.~(\ref{eqn:variance})) will be referred as on-diagonal Partition Noise 
(PN ON), while the off-diagonal ones (second term and part of the third 
term in Eq.~(\ref{eqn:variance})) will be denoted as off-diagonal contribution 
to the partition noise (PN OFF). 

Now let us assume that the number of quantum channel in the source  
is smaller than the one in the drain ($W_S \leq W_D$) and let us 
consider the case of potential barrier wide with respect to the wavelength, 
so that one may neglect tunneling. 
In such a situation, the reflection amplitude matrix $\mathbf{r}$ is equal to 
zero for energies larger than the barrier maximum $E_C$, whereas the 
transmission amplitude matrix is zero for energies 
smaller than $E_C$. 
By means of the unitarity of the scattering matrix 
$\mathbf{s}$, follows $\mathbf{t^{\dagger} t}= \mathbf{I}_S$ for $E>E_C$, 
where $\mathbf{I}_S$ is the identity matrix 
of order $W_S$. Due to reversal time symmetry, there are $W_S$ completely 
opened quantum channels in the drain contact and $W_D - W_S$ completely 
closed. In this situation only the fourth term in Eq.~(\ref{eqn:variance}) 
survives and the noise power spectral density becomes: 
\begin{eqnarray} \label{eqn:openedchannel}
S(0)&=& \frac{2\,e^2 \,W_S}{\pi\hbar}  \int_{E_C}^{+\infty} d\!E   \left[f_S (1\!-\!f_S)+f_D (1\!-\!f_D)\right] \nonumber \\
&=& \frac{2\,e^2 k T \,W_S}{\pi\hbar}\left[f_S(E_C)+f_D(E_C)\right] \,.
\end{eqnarray}   
When $E_{FS}=\, E_{FD}$ such term obviously reduces to the thermal noise 
spectrum $4kTG$, where $G= ( e^2  \,W_S f_S(E_C))/( \pi\hbar) $ is the 
channel conductance at equilibrium.
The fourth term in Eq.~(\ref{eqn:variance}) can be therefore identified with 
the Injection Noise (IN) contribution. 

Equation~(\ref{eqn:variance}) describes correlations between transmitted 
states coming from the same reservoirs [second term in 
Eq.~(\ref{eqn:variance})] 
and between transmitted and reflected states in the source lead 
(third term), with a contribution of opposite sign with respect to the 
first term.
The negative sign derives from Eq.~(\ref{eqn:relation2}), in which 
exchange pairings include a minus sign due to the fermionic 
nature of electrons. 
Note that Eq.~(\ref{eqn:variance}) can be expressed in a symmetric form with 
respect to an exchange between the source and the drain contacts. Indeed, 
by exploiting the unitarity of the scattering matrix, the third term becomes: 
\begin{eqnarray} \label{eqn:term3}
&-&  \! \! \left(\frac{e^2}{\pi \hbar}\right) \!\!\int \!\!d\!E \!\sum_{k \in D} \sum_{p \in S} \!\langle \left[\mathbf{t'^{\dagger}r}\right]_{kp} \left[\mathbf{r^{\dagger}t'}\right]_{pk}  \sigma_{Dk} \sigma_{Sp} \rangle_s  \nonumber \\
&-&  \! \! \left(\frac{e^2}{\pi \hbar}\right) \!\!\int \!\!d\!E \!\sum_{k \in D} \sum_{p \in S} \!\langle \left[\mathbf{r'^{\dagger}t}\right]_{kp} \left[\mathbf{t^{\dagger}r'}\right]_{pk}  \sigma_{Dk} \sigma_{Sp} \rangle_s  \, , 
\end{eqnarray}
which establishes correlations between transmitted and reflected states in the 
source and drain leads. 

Now let us consider the limit when transmission and reflection matrices do not 
depend on random occupation numbers of injected states, i.e. a 
non fluctuating potential profile is 
imposed along the channel. By exploiting the reversal time symmetry ($\mathbf{s}=\,\mathbf{s^t}$, so 
that $\mathbf{t'}=\,\mathbf{t^t}$), the unitarity of the scattering matrix, 
Eq.~(\ref{eqn:variance}) reduces to Landauer-B\"uttiker's 
noise formula~\cite{MBut2}:
\begin{eqnarray} \label{eqn:noiseLB}
S(0)\!\!\! &=& \!\!\!\!\frac{2\,e^2}{\pi\hbar}  \left\{ \int \! d\!E  \sum_{\alpha = S,D} \left(\mathrm{Tr}\left[\mathbf{t^\dagger t} \right]\!-\!\mathrm{Tr}\left[\mathbf{t^\dagger t t^\dagger t} \right] + T_{\alpha} \right)
  f_{\alpha}  \right. \nonumber \\
&-& \! \!\! \int d\!E \, \,\sum_{\alpha = S,D}  T_{\alpha} \, f_{\alpha}^2  \nonumber \\
&-& \!\!\!2  \!\! \int \!d\!E  \left( \mathrm{Tr}\left[\mathbf{t^\dagger t} \right]\!-\!\mathrm{Tr}\left[\mathbf{t^\dagger t t^\dagger t} \right]\right) f_S f_D  \nonumber \\
&+& \left. \! \!\! \!  \! \! \int \! \!d\!E  \sum_{\alpha = S,D}
\left( \mathrm{Tr}\left[\mathbf{t^\dagger t t^\dagger t} \right]\! -\! T_{\alpha} \right) 
\left[f_{\alpha} \left(1\! -\! f_{\alpha}\right)\right] \right\} \, , 
\end{eqnarray} 
where $T_{\alpha}=\sum_{l\neq p \in \alpha}
\left[\mathbf{\tilde{t}} \right]_{\alpha;lp} \left[\mathbf{\tilde{t}} \right]_{\alpha;pl} $ and the sum does not run on the spin. 
Equation~(\ref{eqn:noiseLB}) then reduces to:
\begin{eqnarray} \label{eqn:noiseLB1}
S(0) \!\! &=& \! \!\frac{2\,e^2}{\pi\hbar}\!\! \int \!\!d\!E \left\{ \left[f_S (1\!-\!f_S)+f_D (1\!-\!f_D)\right]\mathrm{Tr}\left[\mathbf{t^\dagger t t^\dagger t} \right]\!+ \right. \nonumber \\
&& \!\!\!\!\!\left.  \left[f_S (1\!-\!f_D)\!+\!f_D (1\!-\!f_S )\right]\! \left( \mathrm{Tr}\left[\mathbf{t^\dagger t} \right]\!-\!\mathrm{Tr}\left[\mathbf{t^\dagger t t^\dagger t}\right]\!\right)\! \right\} \, .\nonumber \\
\end{eqnarray} 
Let us note that Eq.~(\ref{eqn:zerotemperature}) can be recovered as well from 
Eq.~(\ref{eqn:noiseLB1}). Indeed at zero temperature the stochastic injection 
vanishes since random statistics coincides to the Fermi factor. 
In the same way, Eq.~(\ref{eqn:openedchannel}) might be derived from 
Eq.~(\ref{eqn:noiseLB1}), since in this case noise is only due to the 
thermionic emission contribution and fluctuations of the potential profile 
do not play any role in noise. 

\section{Computational methodology and quantitative analysis}

In order to properly include the effect of Coulomb interaction, we 
self-consistently solve the 3D Poisson equation, coupled with the 
Schr\"odinger 
equation with open boundary conditions, within the NEGF formalism, which 
has been implemented in our in-house open 
source simulator {\sl NanoTCAD ViDES}~\cite{ViDES}. For what concerns the 
boundary conditions of Poisson equations, Dirichlet boundary conditions are 
imposed in correspondence of the metal gates, whereas null Neumann 
boundary conditions are applied on the ungated surfaces of the 3D simulation 
domain. 
In particular the 3D Poisson equation reads 
\begin{eqnarray} \label{eqn:Poisson}
\vec{\nabla} \cdot\left[\epsilon \vec{\nabla} \phi\left(\vec{r}\right)\right]= -\left(\rho \left(\vec{r}\right)+\rho_{fix}\left(\vec{r}\right)\right) \, , 
\end{eqnarray} 
where $\phi$ is the electrostatic potential, $\rho_{fix}$ is 
the fixed charge which accounts for ionized impurities in the doped regions, 
and $\rho$ is the charge density per unit volume
\begin{eqnarray} \label{eqn:density}
\rho \left(\vec{r}\right) \!\!\!&=&\!\!\! - e\!\int_{E_i}^{+\infty}\!\! \!d\!E \!\sum_{\alpha= S,D} \sum_{n \in \alpha} DOS_{\alpha n}\left(\vec{r},E\right) \sigma_{\alpha n}(E)  \nonumber \\ 
&+&\!\!\!e \!\!\int_{-\infty}^{E_i} \!\!\!\!\!d\!E \!\!\!\sum_{\alpha= S,D} \sum_{n \in \alpha}\! DOS_{\alpha n}\left(\vec{r},E\right)\! \left[1\!-\!\sigma_{\alpha n}(E)\right] ,
\end{eqnarray} 
where $E_i$ is the mid-gap potential, $DOS_{\alpha n}(\vec{r},E)$ is the 
local density of states associated to channel $n$ injected 
from contact $\alpha$ and $\vec{r}$ is the 3D spatial coordinate.

From a computational point of view, modeling of the stochastic injection 
of electrons from the reservoirs has been performed by means 
of statistical simulations taking into account an ensemble 
of many electron states, i.e. an ensemble of random configurations of injected 
electron states, from both contacts. 
In particular, the whole energy range of integration 
(Eqs.~(\ref{eqn:variance}) and~(\ref{eqn:density})) has been 
uniformly discretized with energy step $\Delta E$. 
Then, in order to obtain a random injection configuration, 
a random number $ r $ uniformly distributed between 
0 and 1 has been extracted for each electron state represented by energy $E$, 
reservoir $\alpha$ and quantum channel $n$~\cite{ABetti}.
More in detail, the state is occupied if $r$ is smaller than the Fermi-Dirac 
factor, i.e. $\sigma_{S n}(E)$ [$\sigma_{D n}(E)$] is 1 if 
$ r < f_S(E)\, [f_D(E)]$, and 0 otherwise. 

The random injection configuration generated in this way has been 
then inserted in Eq.~(\ref{eqn:density}) and self-consistent 
solution of Eqs.~(\ref{eqn:Poisson}) and~(\ref{eqn:density}) and the 
Schr\"odinger equations has been performed. 
Once convergence has been reached, the transmission 
($\mathbf{t},\mathbf{t'}$) and reflection ($\mathbf{r}$) matrices are 
computed. The procedure is repeated several times in order to gather data 
from a reasonable ensemble. 
In our case, we have verified that an ensemble of 
500 random configurations represents a good trade-off between 
computational cost and accuracy. 
Finally, the power spectral density $S(0)$ has been extracted by means 
of Eq.~(\ref{eqn:variance}). 

In the following, we will refer to self-consistent Monte Carlo 
simulations (SC-MC), when statistical simulations using the procedure 
described above, i.e. inserting random occupations 
$\sigma_{Sn}(E)$ and $\sigma_{Dn}(E)$ in Eq.~(\ref{eqn:density}), 
are performed. 
Instead we will refer  to self-consistent (SC) simulations 
when the Poisson-Schr\"odinger equations are solved considering 
$f_S$ and $f_D$ in Eq.~(\ref{eqn:density}). 
SC-MC simulations of randomly injected electrons allow to consider both the 
effect of Pauli and Coulomb interaction on noise. 

\begin{figure} [tbp]
\begin{center}
\includegraphics[width=8.6cm]{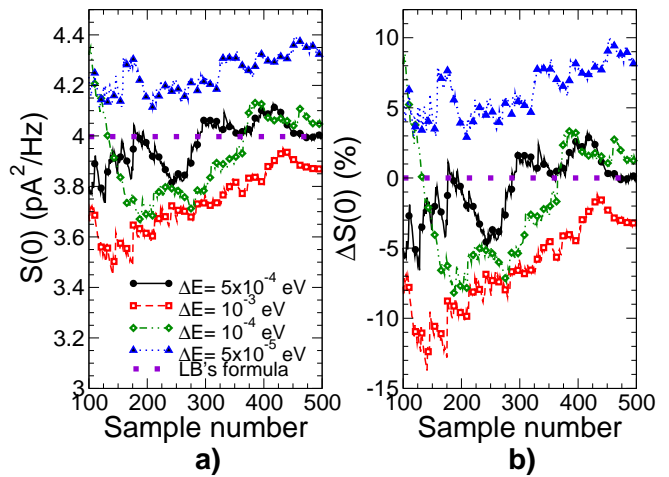}
\end{center}
\caption{(Color online) a) Noise power spectral density $S(0)$ obtained from 
Eq.~(\ref{eqn:variance}) for a given potential 
as a function of current 
sample number for four different energy steps. b) Relative deviation of 
$S(0)$ with respect to Landauer-B\"uttiker's limit~(\ref{eqn:noiseLB1}). 
The simulated structure is the SNW-FET shown in Fig.~\ref{fig:struttura}.}
\label{fig:sceltadESNWVg0_5}
\end{figure}

From a numerical point of view, particular 
attention has to be posed on the choice of the energy step $ \Delta E $. 
In Fig.~\ref{fig:sceltadESNWVg0_5} the noise power spectrum computed 
by keeping fixed the potential profile 
along the channel and performing statistical Monte Carlo simulations of 
randomly injected electrons is shown for four energy steps. 
As already proved in Eq.~(\ref{eqn:noiseLB}), the convergence to 
Landauer-B\"uttiker's limit is ensured for all the considered energy steps: 
as can be seen, $\Delta E= $ 5 $\times$10$^{-4}$ eV provides faster 
convergence as 
compared to the other values with a relative error close to 0.16\%.

Let us point out that the NEGF formalism computes directly the total Green's 
function $\mathbf{G}$ of the channel and the broadening function of the source 
($\mathbf{\Gamma}_S$) and drain ($\mathbf{\Gamma}_D$) leads, rather than the 
scattering matrix $\mathbf{s}$, that relates the outgoing waves amplitudes 
to the incoming waves amplitudes at different reservoirs. 
In order to obtain the matrix $\mathbf{s}$, we have exploited the 
\begin{em} Fisher-Lee relation \end{em}~\cite{Fisher}, which expresses the 
elements of the $\mathbf{s}$-matrix in terms of the Green's function 
$\mathbf{G}$ and transverse mode eigenfunctions 
(see Appendix~\ref{Appendix}). 

\section{Results}

\begin{figure} [tbp]
\begin{center}
\includegraphics[width=8.6cm]{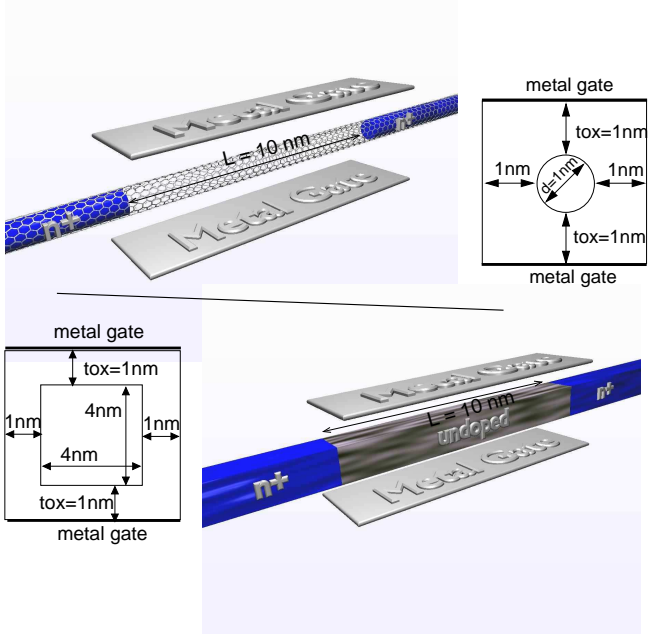}
\end{center}
\vspace{-0.5cm}
\caption{(Color online) 3-D structures and transversal cross sections of the simulated CNT 
(top) and SNW-FETs (bottom).}
\label{fig:struttura}
\end{figure}

The approach described in the previous section has been 
used to study the behavior of shot noise in quasi-1D channel of 
CNT-FETs and SNW-FETs with identical reservoirs (Fig.~\ref{fig:struttura}). 
We consider a (13,0) CNT embedded in SiO$_{2}$ with 
oxide thickness equal to 1 nm, an 
undoped channel of 10 nm and n-doped CNT extensions 10 nm long, with a molar 
fraction $f=\,5 \times 10^{-3}$. 
The SNW-FET has an oxide thickness ($t_{ox}$) equal to 1 nm and the 
channel length ($L$) is 10 nm. The channel is undoped and the source and drain 
extensions (10 nm long) are doped with $ N_D = \, 10^{20} $ cm$^{-3}$. The 
device cross section is 4$\times$4~nm$^2$.

From a numerical point of view, a p$_z$-orbital tight-binding Hamiltonian 
has been assumed for CNTs~\cite{GFiori2,JGuo2}, whereas an effective mass 
approximation has been considered for SNWs~\cite{GFiori1,JWang} by means of 
an adiabatic decoupling in a set of
two-dimensional equations in the transversal plane and in a set of 
one-dimensional equations in the longitudinal direction for each 1D subband. 
For both devices, we have 
developed a quantum ballistic transport model with semi-infinite 
extensions at their ends. 
A mode space approach has been adopted, since only 
the lowest subbands take part to transport. In particular, we have 
verified that four modes are enough to compute the mean current 
both in the ohmic and saturation regions. 
All calculations have been performed at room temperature ($T=$ 300 K). 

\begin{figure} [tbp]
\begin{center}
\includegraphics[width=8.6cm]{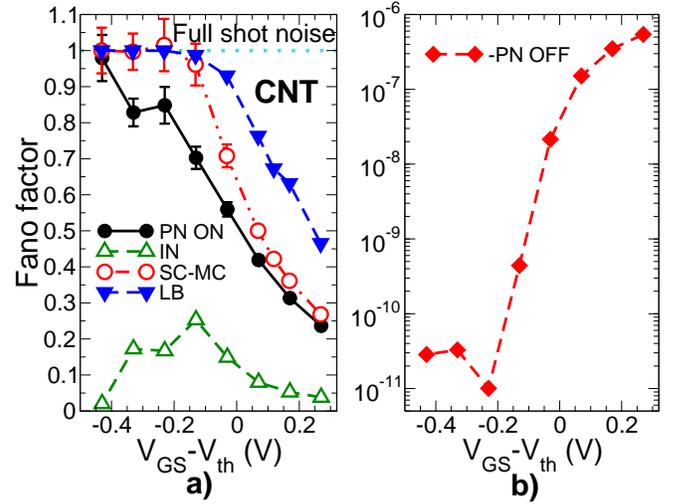}
\end{center}
\caption{(Color online) Contributions to the Fano factor in a CNT-FET of the on-diagonal and 
off-diagonal partition noise and of the injection noise 
(respectively on-diagonal and off-diagonal part of the 
first three terms, and fourth term in Eq.~(\ref{eqn:variance})) as a function
of the gate overdrive  $V_{GS}-V_{th}$ for a drain-to-source bias 
$V_{DS} =$ 0.5~V. 
a) The on-diagonal partition (PN ON, solid circles), the injection 
(IN, open triangles up) 
and the full noise (open circles) computed by means of SC-MC simulations 
are shown. 
The Fano factor computed by exploiting Landauer-B\"uttiker's formula~(\ref{eqn:noiseLB1}) and SC simulations (solid triangles down) is also shown. 
b) Off-diagonal partition noise contribution (PN OFF) to $F$ due to 
correlation between transmitted states and between transmitted 
and reflected states.}
\label{fig:fig1}
\end{figure}

\begin{figure} [h]
\vspace{-0.2cm}
\begin{center}
\includegraphics[width=8.6cm]{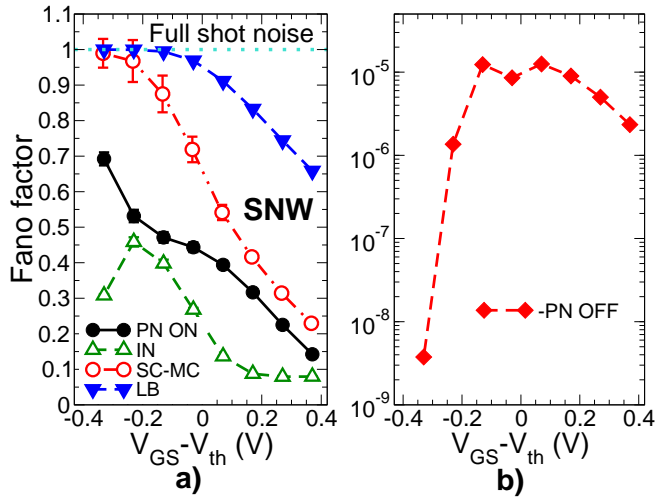}
\end{center}
\caption{(Color online) Contributions to the Fano factor in a SNW-FET of the on-diagonal 
and off-diagonal partition noise and of the injection noise, 
obtained for $V_{DS} =$ 0.5~V, as a function
of the gate overdrive $V_{GS}-V_{th}$ in a SNW-FET. 
In a) the on-diagonal partition, the injection and the full noise computed 
by means of SC-MC simulations (both Pauli and Coulomb interactions taken 
into account)
are shown together with results obtained by means of Eq.~(\ref{eqn:noiseLB1}). 
b) Off-diagonal partition noise due to correlation between transmitted states 
and between transmitted and reflected states.}
\label{fig:fig2}
\end{figure}

Let us focus our attention on the Fano factor $F$, defined as the ratio 
of the actual noise power spectrum $S(0)$ to the full shot noise 
$2 q \langle I \rangle $. 
In Figs.~\ref{fig:fig1} and~\ref{fig:fig2} the contributions to $F$ of 
partition noise (first three terms in Eq.~(\ref{eqn:variance})) and injection 
noise (fourth term in Eq.~(\ref{eqn:variance}))  are shown, 
as a function of the gate 
overdrive $V_{GS}-V_{th}$ for a drain-to-source bias $V_{DS}=$ 0.5~V for 
CNT-FETs and SNW-FETs, respectively: results have been obtained by 
means of SC-MC simulations. 
The threshold voltage $V_{th}$ at $V_{DS}=$ 0.5~V is 0.43~V for the CNT-FET 
and 0.13~V for the SNW-FET. 
In particular, Figs.~\ref{fig:fig1}a and~\ref{fig:fig2}a refer to the 
on-diagonal contribution to the partition noise (solid circles), to 
the injection noise (open triangles up) and to the 
complete Fano factor (open circles) obtained by means of Eq.~(\ref{eqn:variance}), i.e. Pauli and Coulomb interactions simultaneously 
considered. 
We present also the Fano factor (solid triangles down) computed by 
applying Eq.~(\ref{eqn:noiseLB1}) on the self-consistent potential profile, 
i.e. when only Pauli exclusion principle is included. 
In Figs.~\ref{fig:fig1}b and~\ref{fig:fig2}b we show the contribution 
of the off-diagonal partition noise to $F$, which provides a measure of 
mode-mixing and of exchange interference effects.  

As can be seen in Fig.~\ref{fig:fig1}a and Fig.~\ref{fig:fig2}a, in the 
sub-threshold regime ($V_{GS} - V_{th}< $ -0.2~V, 
$\langle I \rangle <$ 10$^{-9}$ A) 
the Poissonian noise for a non-degenerate injection is recovered, since 
electron-electron interactions are negligible due to the very small amount 
of mobile charge in the channel.
In the strong inversion regime instead ($V_{GS} - V_{th}> $~0~V, 
$\langle I \rangle>$ 10$^{-6}$ A), 
noise is greatly suppressed with respect to the full shot value.
In particular for a SNW-FET, at $ V_{GS} - V_{th}\approx$~0.4~V 
($\langle I \rangle\approx$ 2.4 $\times $ 10$^{-5}$ A), combined Pauli and 
Coulomb interactions 
suppress shot noise down to 22\% of the full shot noise value, 
while for CNT-FET the 
Fano factor is equal to 0.27 at $ V_{GS} - V_{th}\approx$ 0.3~V 
($\langle I \rangle\approx$ 1.4 $\times $ 10$^{-5}$ A). 
This is due to the fact that as soon as an electron is injected, the 
barrier height along the channel increases, leading to a reduced transmission 
probability for other electrons. 

As shown in Fig.~\ref{fig:fig1}a, the dominant noise source in 
ballistic CNT-FETs is the on-diagonal partition noise and the noise due to 
the intrinsic thermal agitations 
of charge carriers in the contacts (injection noise), which 
is at most the 36~\% of the partition noise 
($V_{GS}-V_{th}\approx-0.1$~V). 
Nearly identical results are shown for SNW-FETs, with the 
exception of a stronger contribution given by the injection noise, up to 
the 86~\% of the on-diagonal partition term ($V_{GS}-V_{th}\approx-0.2$~V). 
Moreover, the behavior of the two noise components, as a function 
of $V_{GS}-V_{th}$, is very similar for both CNT- and SNW-FETs: $F$ tends to 1 
in the subthreshold regime, while in strong inversion regime shot noise is 
strongly suppressed.

Let us stress that an SC-MC simulation exploiting Eq.~(\ref{eqn:variance}) 
is mandatory for a quantitative evaluation of noise. 
Indeed, by only considering Pauli exclusion principle 
through formula~(\ref{eqn:noiseLB1}), one would have overestimated 
shot noise by 180~\% for SNW-FET ($V_{GS}-V_{th}\approx$ 0.4~V) and 
by 70~\% for CNT-FET ($V_{GS}-V_{th}\approx$ 0.3~V)~\cite{ABetti,ABettiTED}. 

It is interesting to observe that the off-diagonal contribution 
to partition noise, due to exchange correlations between 
transmitted states and between transmitted and reflected states, has a strong 
dependence on the height of the potential profile along the channel 
(variation of 5 orders of magnitude for CNT-FETs) and is negligible for quasi 
one-dimensional FETs. 
In particular, for CNT-FETs such term is at most 5 orders of magnitude 
smaller than the on-diagonal partition noise or injection noise in the 
strong inversion regime 
($V_{GS}-V_{th}\approx$ 0.3~V), while in the subthreshold regime its magnitude 
still reduces (about 10$^{-11}$ for $V_{GS}-V_{th}\approx$ -0.4~V). 
For SNW-FETs we have obtained similar results: the off-diagonal partition 
noise is indeed at most 5 orders of magnitude smaller than the 
other two contributions. 

In such conditions, transmission occurs only along separate quantum channels 
and an uncoupled mode approach is also accurate. Indeed, off-diagonal 
partition noise provides an interesting information on the strength of 
the mode-coupling which, 
as already seen, is very small. 
In particular, neglecting this term, results obtained from 
Eq.~(\ref{eqn:variance}) can be recovered as well. 

In the previous discussion, carriers from different quantum channels do not 
interfere. However, since 
we deal with a many indistinguishable particle system, such effects 
can come into play. 
To this purpose, we investigate in more detail two examples in which 
exchange pairings, that include also exchange interference effects, 
give a non negligible contribution to drain current noise. 
In the past exchange interference effects have been already predicted for 
example in ballistic conductor with an elastic scattering center in the 
channel~\cite{Gramespacher}, in diffusive four-terminal conductors of 
arbitrary shape~\cite{Sukhorukov} and in 
quantum dot in the quantum Hall regime~\cite{ButtikerPRL}, connected to two 
leads via quantum point contacts.

\begin{figure} [tbp]
\begin{center}
\includegraphics[width=8.6cm]{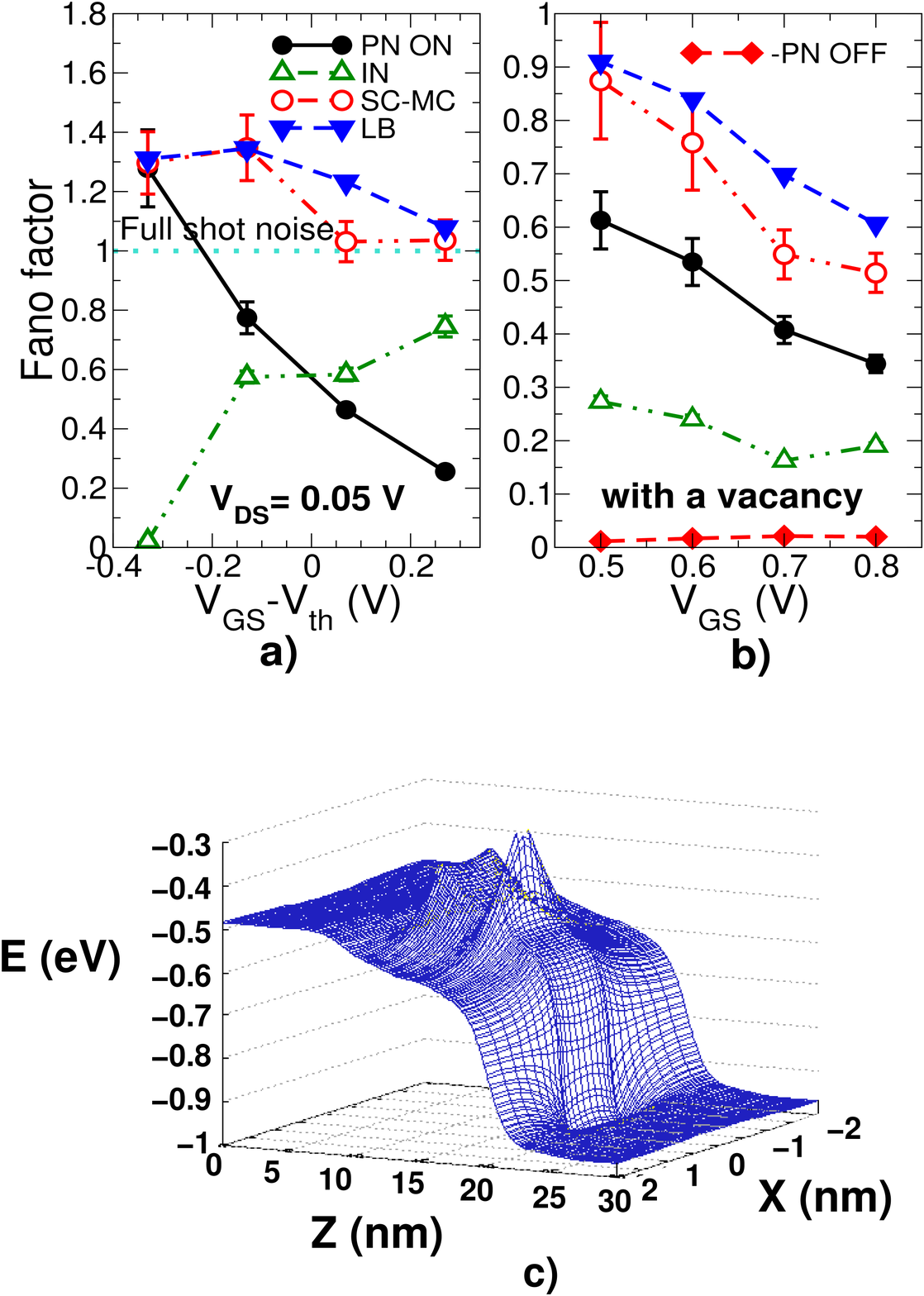}
\end{center}
\caption{(Color online) a): Contributions to the Fano factor $F$ by the on-diagonal partition 
noise (solid circles), 
and the injection noise (open triangles up) as a function of the gate 
overdrive $V_{GS}-V_{th}$, for a drain-to-source bias 
$V_{DS} =$ 50~mV. The simulated device is a CNT-FET. 
The full noise computed by means of SC-MC simulations 
(open circles, both Pauli and Coulomb interactions taken into account) and 
applying Eq.~(\ref{eqn:noiseLB1}) 
(solid triangles down, only 
Pauli exclusion considered) is also shown. b): Contributions to $F$ by 
the on-diagonal and off-diagonal partition noise and by the injection noise 
(exploiting Eq.~(\ref{eqn:variance})) as a function of the gate overdrive 
for a CNT-FET 
with a vacancy in a site at the center of the channel. The drain-to-source 
bias is 0.5 V. c): Self-consistent midgap potential obtained by using the 
Fermi statistics for a gate voltage $V_{GS}=$ 0.7 V and a 
bias $V_{DS}=$ 0.5 V. $Z$ is the transport direction along the channel, 
$X$ is a transversal direction. The simulated device is the same of b).}
\label{fig:fig4} 
\end{figure}

In the first case we discuss, mode-mixing does not appear, i.e. the 
non-diagonal elements of the matrices 
$\mathbf{t^{\dagger}t}$ and $\mathbf{t'^{\dagger}r}$ are negligible 
with respect to the diagonal ones. 
Since the off-diagonal partition noise is negligible and since in the 
third term in Eq.~(\ref{eqn:variance}) only contributions with indices 
$l=n=k=p$ survive, exchange interference effects 
do not contribute to electrical noise. 
We consider a CNT-FET at low bias condition: $V_{DS}=$ 50 mV. 
In Fig.~\ref{fig:fig4}a the on-diagonal partition noise, the injection noise 
and correlations due to the off-diagonal partition noise, evaluated performing 
statistical SC-MC simulations, are shown. 
In this case, on-diagonal correlations between transmitted and reflected 
states in the source lead (in the same quantum channel) extremely affect noise. 
Indeed, at the energies at which reflection events in the source lead are 
allowed, also electrons coming from D can be transmitted into the injecting 
contact $S$, since the corresponding energy states in $D$ are occupied and 
the barrier height is small. 
Instead the exchange correlations represented by the off-diagonal partition 
noise are negligible, since they are at least 5 order of 
magnitude smaller than the other three terms in Eq.~(\ref{eqn:variance}).     
Note that the noise enhancement obtained both in the inversion and 
subthreshold regimes is due to the fact that at low bias the current 
$\langle I \rangle$ becomes small, while the noise power spectrum $S(0)$ 
tends to a finite value, because of the thermal noise contribution.

Let now consider the situation in which modes are coupled and 
exchange interference effects, through the off-diagonal partition noise, 
contribute to drain current fluctuations. 
We consider the interesting case 
in which a vacancy, i.e. a missing carbon atom, is placed at the center 
of the channel of a (13,0) CNT-FET. 
From a numerical point of view, this defect can be modelled by introducing 
a strong repulsive potential (i.e.~+8 eV, much larger than the energy gap of 
a (13,0) CNT: $E_{gap}\approx$ 0.75 eV) in correspondence of 
such site, thus acting as a barrier for transmission in the middle of the 
channel (Fig.~\ref{fig:fig4}c). 

In Fig.~\ref{fig:fig4}b the three noise sources in Eq.~(\ref{eqn:variance}) 
(on- and off-diagonal partition noise, injection noise) are plotted as a 
function of the gate voltage $V_{GS}$ in the above threshold regime 
for $V_{DS}=$ 0.5~V, along with 
the full Fano factor computed performing SC and SC-MC simulations. 
Remarkably, in this case a mode space approach taking into account all 
modes (i.e. 13) is mandatory in order to reproduce all correlation 
effects on noise. 
As can be seen, off-diagonal exchange correlations gives
rise to a not negligible correction to the Fano factor 
($\approx$ 4~\% of the full Fano factor at $V_{GS}=$ 0.8~V). 
We observe that such correlations are only established between transmitted 
electrons states (second term in Eq.~(\ref{eqn:variance})), while 
correlations between reflected and transmitted electron states 
(third term in Eq.~(\ref{eqn:variance}))  
are negligible since almost all electrons injected from the receiving 
contact $D$ are reflected back because of the high bias condition. 
In this paper we have assumed phase-coherent quantum transport at room 
temperature. Our tools cannot include electron-phonon interaction, that 
a room temperature may play a role even in nanoscale devices. 
Ref.~\cite{HPark} has considered the effect of electron-phonon scattering 
and has neglected Coulomb interaction: they find that electron-phonon 
scattering increase shot noise in the above threshold regime, due to the 
broadening of the energy range of electron states contributing to transport.

\section{Conclusion}

We have developed a novel and general approach to study 
shot noise in ballistic quasi one-dimensional CNT-FETs and 
SNW-FETs. 
By means of a statistical approach within the second quantization formalism, 
we have shown that the 
Landauer-B\"uttiker noise formula (Eq.~(\ref{eqn:noiseLB1})) can 
be generalized 
to include also Coulomb repulsion among electrons. 
This point is crucial, since we have verified that by only using 
Landauer-B\"uttiker noise formula, i.e. considering only Pauli exclusion 
principle, one can overestimate shot noise by as much as 180~\%. 

From a computational point of view, we have quantitatively evaluated shot 
noise in CNT-FETs and SNW-FETs by 
self-consistently solving the electrostatics and the transport 
equations  within the NEGF formalism, for a large ensemble of snapshots of device operation, 
each corresponding to a different configuration of the occupation of injected states.

Furthermore, with our approach we are able to observe a rectification of the DC characteristics due to 
fluctuations of the channel potential, and to identify and evaluate quantitatively the different contributions
to shot noise. We are also able to consider the exchange interference effects, which are often negligible
but can be measurable when a defect, introducing significant mode mixing, is inserted in the channel. 

\begin{acknowledgments}
The work was supported in part by the EC Seventh 
Framework Program under the Network of Excellence NANOSIL (Contract 216171), 
and by the European Science Foundation EUROCORES Program Fundamentals of 
Nanoelectronics, through funds from CNR and the EC Sixth Framework Program, 
under project DEWINT (ContractERAS-CT-2003-980409). The authors would like 
to thank Prof. M. B\"uttiker for fruitful discussion. 
\end{acknowledgments}

\appendix*
\section{} \label{Appendix}

Let us consider a 2D channel of length $L$ and denote with $x$ and $y$ 
the longitudinal direction and the transverse one, respectively. 
If the interface between the lead S (D) and the conductor is defined by 
$ x_S=0$ 
($x_D=0 $), $ \mathbf{G}_{D S}\left(y_D;y_S\right)= \mathbf{G}_{D S}\left(x_D=0,y_D;x_S=0,y_S\right)$ represents the wavefunction at $\left(x_D=0,y_D\right)$ due to an excitation at $\left(x_S=0,y_S\right)$.
In real space the \begin {em}Fisher-Lee relation\end{em} reads:
\begin{eqnarray} \label{eqn:Fisher} 
\mathbf{s}_{n m}\!\!\!\!&=&\!\!\!\! -\delta_{nm} + \frac {i \hbar \sqrt{v_n v_m}}{a}\!\! \int \!\!dy_D \!\!\int \!\!dy_S \, \chi_n \left(y_D\right) \mathbf{G}_{D S}\left(y_D;y_S\right) \nonumber \\ 
&&\chi_m \left(y_S\right) 
\end{eqnarray}
where \begin{math} n \end {math} is a mode outgoing at lead D with velocity 
$ v_n$, $ m$ is a mode incoming at lead S with velocity $ v_m$ and $a$ is 
the lattice constant along the $ x $ direction.
In the {\bf k}-representation, for a conductor of uniform 
cross-section, we can exploit a mode representation in the transverse 
direction and a plane wave representation in the longitudinal direction and 
(\ref{eqn:Fisher}) becomes:
\begin{eqnarray} \label{eqn:FisherK} 
\mathbf{s}_{n m} =  
-\delta_{nm} +\frac{ i \hbar \sqrt{v_n v_m}}{L} \, \, \mathbf{G}_{D S}\left(n, m\right) 
\end{eqnarray}
where $ \mathbf{G}_{D S}\left(n, m\right) = \mathbf{G}_{D S}\left(n,k_n;m,k_m\right) $ and $ k_n $ is the 
longitudinal wavevector of the transverse mode $n$. 
Let us assume both leads to be identical and denote with $\{ k_1^S ,..., k_{N}^S\}$ ($ \{ k_1^D ,...,k_{N}^D \} $) the set of wavevectors 
associated to the $N$ modes coming from the 
lead S (D). 
Since the only non-zero components of the self-energy involve the end-points, 
in the {\bf k}-representation $\mathbf{\Gamma}_S$ and $\mathbf{\Gamma}_D$ can be expressed as 
$$
\begin{array}{cc}
\mathbf{\Gamma}_S = \left(
\begin{array}{cc}
\mathbf{\Gamma}_{S;11} & \mathbf{0}  \\
 \mathbf{0} & \mathbf{0}  \\
\end{array}
\right)_{2N \times 2N} \, &
\,  \mathbf{\Gamma}_D = \left(
\begin{array}{cc}
\mathbf{0} & \mathbf{0}  \\
\mathbf{0}  & \mathbf{\Gamma}_{D;22}  \\
\end{array}
\right)_{2N \times 2N} \\
\end{array}
$$
where $ \mathbf{\Gamma}_{S;11}\left(n,m\right)=\delta_{nm}\frac{\hbar v (k_n^S)}{L}$ $\forall \, n,m \in S$ and $ \mathbf{\Gamma}_{D;22}\left(n,m\right)=\delta_{nm}\frac{\hbar v (k_n^D)}{L}$    $\forall \, n,m \in D$.

Generalization to a CNT-FET structure is straightforward. Let us 
indicate with $N_C$ and $N_M$ the number of carbon atoms rings and the number 
of modes propagating along the channel, respectively. Since the coupling 
between the identical reservoirs and the channel involve only the 
end-rings of the channel, 
$\mathbf{\Gamma}_S$ 
and $\mathbf{\Gamma}_D$ are $(N_M N_C)\times (N_M N_C)$ 
diagonal matrix and the only non-zero blocks are the first one and 
the latter one, respectively: 
\begin{eqnarray} \label{eqn:gamma}
\mathbf{\Gamma}_{S;11}\left(n,m\right)=\delta_{nm}\frac{\hbar v (k_n)}{L} \,\,\, \,\forall n,m = 1,\cdots,N_M \nonumber \\
\mathbf{\Gamma}_{D;N_c N_c}\left(n,m\right)=\delta_{nm}\frac{\hbar v (k_n)}{L} \,\,\,\, \forall n,m = 1,\cdots,N_M  
\end{eqnarray}
By exploiting Eqs.~(\ref{eqn:FisherK}) and~(\ref{eqn:gamma}) 
we can find the 
transmission ($\mathbf{t}$) and reflection ($\mathbf{r}$) amplitude matrix:
\begin{eqnarray} \label{eqn:tr}
\mathbf{t}_{nm}= i \sqrt{\mathbf{\Gamma}_{D;N_C N_C} \left(n,n \right)} \mathbf{G}_{N_C 1}\left(n,m\right) \sqrt{\mathbf{\Gamma}_{S;1 1} \left(m,m\right)}  \nonumber \\
\mathbf{r}_{nm}\!=\! -\delta_{nm} + i \sqrt{\mathbf{\Gamma}_{S;1 1}\left(n,n \right)} \mathbf{G}_{1 1}\left(n,m\right) \! \sqrt{\mathbf{\Gamma}_{S;1 1} \left(m,m\right)}  \nonumber \\
\end{eqnarray}
Since at zero magnetic field $\mathbf{t'}= \mathbf{t^t}$, 
relations~(\ref{eqn:tr}) is all we need to compute the power spectral 
density~(\ref{eqn:noisepower}) from Eq.~(\ref{eqn:variance}). 
A similar procedure has been adopted for SNW-FETs where, from a computational 
point of view, the channel has been discretized in a sequence of slices in the 
longitudinal direction. In this case Eqs. in~(\ref{eqn:tr}) are 
obtained as well, but replacing the number of rings with the number of slices. 


\end{document}